\documentclass{article}
\usepackage{authblk}
\usepackage{amsmath}

\begin{document}

\title{On the gauge invariance of the locally averaged Friedmann Universe and the Hubble tension}
\author{Masanori Tomonaga \thanks{i2185088@cc.kyoto-su.ac.jp}}
\author{Toshifumi Futamase \thanks{tof@cc.kyoto-su.ac.jp}}

\affil{Faculty of Science, Kyoto Sangyo University, Motoyama, Kamigamo, Kita-ku, Kyoto, 603-8555 Japan}
\date{}
\maketitle

\begin{abstract}
    The Hubble tension cast a blight on the standard cosmology.
    As a possible attitude to the problem, the local variation of the expansion rate in an inhomogeneous cosmology has been proposed where the spatial averaging over a finite domain was introduced in order to construct local Friedmann spacetime. 
    However, it is not clear that the concept of the spatial averaging itself is gauge invariant or not. 
    Namely the conclusion obtained by the averaging in a particular gauge is physically meaningful or not. 
    In this paper we address this question, namely the gauge invariance of the spatial averaging in cosmology . 
    We show that the answer is positive by studying the spatial average in the gauge-invariant cosmological perturbation theory.
\end{abstract}

\section{Introduction}
The Hubble constant $H_0$ is one of the most important cosmological parameter since it characterizes the global properties of our universe.
The standard cosmology is based on the assumption of the homogeneity and isotropy. 
Thus the Hubble parameter $H_0$ is regarded as a constant over at least the horizon scale which is also the prediction of the inflationary scenario.
However recent observations suggest a non negligible difference between local and global (or recent and old) Hubble parameter.\cite{2020,riess20162} \\
There has been a large number of studies which try to resolve the discrepancy.\cite{ichiki2016relationship,bolejko2018emerging,tomita2017cosmological,tomita2018super,benetti2019looking,akarsu2019constraints,cea2022ellipsoidal,blinov2020warm,bento2002generalized,2021realm,2021darksector,berghaus2020thermal,jedamzik2020relieving,macpherson2018trouble,cai2021chameleon,martin2021hubble}
We regard that  the difference of the local and global Hubble parameter is real and be
explained  by the inhomogeneous distribution of the matter. 
In fact the observation of the K-band luminosity density seems to suggest that region with several hundred Mpc
around us has low density with density contrast $\delta_K\sim -0.5$ compared with
the globally averaged density.\cite{keenan2013evidence} Furthermore there is some indication that the voids are actually low density by weak lensing observation.
Thus it will be meaningful to pursue the indication of the cosmological inhomogeneity.

The homogeneous and isotropic universe(here we call Friedmann Universe)  appears as
the  result of some kind of averaging procedure since the universe is actually very inhomogeneous. 
There are various ways to averaging inhomogeneous universe (such as the light-corn averaging that is directly related with observational quantities). In this paper, we only consider the scalar perturbations in the linear order and the spatial averaging.
\cite{buchert2002regional,wiltshire2007cosmic,buchert1995averaging,ben2012backreaction,gasperini2011light,fanizza2020generalized,yoo2017gauge,korzynski2010covariant,zalaletdinov2004spacetime,clarkson2012observational,coley2006averaging,heinesen2019covariance,buchert2001average,buchert2000average,boersma1998averaging}
Our purpose is not studying the averaging itself in general inhomogeneous spacetimes, but rather  the gauge dependence of the relationship between locally averaged and globally averaged spacetime in the linearly perturbed universe using the spatial averaging.
By adopting the spatial averaging defined below, we were able to derive a locally averaged Friedmann universe and have obtained the following  relation between the locally average Hubble parameter and the globally averaged Hubble parameter within the framework of the general relativistic perturbation theory\cite{kasai2019possible}
\begin{equation}
    H_{D0} = H_0\left(1 - \frac{1}{3}\langle\delta\rangle_{Dt_0}\right)
\end{equation}
where $H_{D0}$ is the averaged Hubble parameter at the present time $t_0$ over a finite domain $D$, and $H_0$ is the global, or the horizon scale Hubble parameter, and $\langle\varDelta\rangle_{Dt_0}$ is the present density contrast average over the domain $D$.
However the treatment is carried out in the comoving synchronous and Newtonian gauge, and there is some question if the averaging and the result are gauge invariant or not.

In order to answer the question we study the spatial averaging in the framework of the gauge invariant cosmological perturbation theory, and find that the averaging has a gauge invariant meaning and thus the results obtained by the averaging has physically relevant.

\section{3+1 decomposition of Einstein equations and the spatial averaging}
We follow the previous treatment, namely 3 + 1 decomposition of Einstein equations.
We use the following convention:Greek indices $\mu,\nu,\cdots$run from $0\sim 3$, Latin indices $i,j,\cdots$run from $1\sim 3$, and the speed of light is unity, $c=1$.

We write the line element as follows.
\begin{equation}
    ds^2 = -N^2dt^2 + q_{ij}\left(dx^i + N^idt\right)\left(dx^j + N^jdt\right)
\end{equation}
Then we obtained the $3+1$ decomposition of the Einstein equations as follows.
\begin{align}
    \left(K^i_i\right)^2 - K^i_jK^j_i + ~^{(3)}R^i_i &= 16\pi G \frac{T_{00}}{N^2} \label{00} \\
    K^j_{i|j} - K^j_{j|i} &= 8\pi G\frac{T_{0i}}{N} \label{0i} \\
    \mathcal{L}_nK_{ij} = 2K_{il}K^l_j - K^l_lK_{ij} - ~^{(3)}R_{ij} &+ N_{|ij} + 8\pi G\left(T_{ij} - \frac{1}{2}q_{ij}T\right) \label{ij}
\end{align}
where $\mathcal{L}_n$ is Lie derivative, $|$ denotes the three-dimensional derivative with respect to $g_{ij}$.
$~^{(3)}R^i_j$ is the three-dimensional Ricci tensor and $K^i_j$ is the extrinsic curvature defined as follows.
\begin{equation}
    K_{ij} \equiv \nabla_{i}n_j = \frac{1}{2N}\left(\frac{\partial q_{ij}}{\partial t}  - N_{i|j} - N_{j|i}\right)
\end{equation}
$n_{\mu}(\text{or}~n^{\mu})$ is the normal unit vector at t=const. hypersurface, defined as
\begin{equation}
    n_{\mu} = -N\nabla_{\mu}t,\quad n^{\mu} = N^{-1}\left[\left(\partial_t\right)^{\mu} - N^{\mu}\right]
\end{equation}
We consider the dust as our fluid and thus the energy momentum tensor is
\begin{equation}
    T^{\mu\nu} = \rho u^{\mu} u^{\nu}
\end{equation}
where $u^{\mu}$ is the four velocity.

We average the above equations \eqref{00},\eqref{0i},\eqref{ij} over sufficiently large
region $D$ where sufficient number of galaxies are included and smaller than the horizon scale $H$. The averaging of the quantity $Q$ over the region $D$ is defined as  follows.
\begin{equation}
\langle Q \rangle_D\equiv \frac{1}{V_D}\int_D \sqrt{\gamma} Qd^3x
\end{equation}
with the volume defined as
\begin{equation}
V_D\equiv \int_D\sqrt{\gamma}d^3x
\end{equation}

Since the observation of the Cosmic Microwave Background strongly suggests that our universe is homogeneous and isotropic at least over the horizon scale,
thus it is natural to assume that the horizon scale averaging satisfies $\langle N\rangle_H=1, \, \langle N^i\rangle_H=0$ and  $\langle q_{ij}\rangle_H=a^2(t)\gamma_{ij}$ where $\gamma_{ij}$ is the 3 metric of a  constant curvature space, as well as ${\langle \rho\rangle}_H={\bar \rho}, \, \langle u^0=1\rangle_H=1$ and $\langle u^i\rangle_H=0$. This give us the usual background geometry.
\begin{equation}
    ds^2 = -dt^2 + a^2\gamma_{ij}dx^idx^j
\end{equation}

With this background, the extrinsic and three-dimensional curvature are given by
\begin{gather}
    K^i_j = \frac{\dot{a}}{a}\delta^i_j \\
    R^i_j = \frac{2K}{a^2}\delta^i_j
\end{gather}
where $K$ is a signature of the spatial curvature. As easily confirmed, the above choices
made that equation (\ref{0i})  satisfies trivially and
simplify (\ref{00}) and (\ref{ij}) as follows.
\begin{gather}
    \left(\frac{\dot{a}}{a}\right)^2 + \frac{K}{a^2} = \frac{8\pi G}{3}{\bar \rho} \\
    \dot{{\bar \rho}} + 3\frac{\dot{a}}{a}{\bar \rho} = 0
\end{gather}
In the following we assume $K = 0$ for simplicity.

\section{Guage invariant linear perturbation theory}
The universe is well described by a Friedmann universe on the horizon scale as described above. 
This geometry should be regarded as the result of some kind of averaging since the universe is very inhomogeneous in local scales.  
Therefore it is natural to expect the averaged Friedmann universe may not be the same if the averaged region is different.

  Although the density contrast indicated from the K band luminosity density is of the order of $-0.5$, the metric perturbation may not be so large
when the local scale is sufficiently smaller than the horizon scale since we have the relation between the density contrast and the Newtonian potential as $\Delta\Phi\sim G{\bar \rho}\delta\sim H^2_0\delta$. 
Therefore we assume that the metric fluctuations  are small and we use with linear perturbation theory. 
Unfortunately this means we restrict the situation where the density contrast is also small enough.

Then we write the metric and the energy density in the perturbed universe as follows:
\begin{align}
    ds^2 &= -\left(1 + 2A\right)dt^2 + a^2\left(\delta_{ij} + 2E_{,ij} + 2F\delta_{ij}\right)\left(dx^i + B^{,i}dt\right)\left(dx^j + B^{,j}dt\right) \\
    \rho &= \rho_b\left(1 + \delta\right)
\end{align}
where we consider only scalar perturbation and the scalars $A, E, F$ and $B$ as well as $\delta$ are small quantities.

In this metric, the proper time can be defined as $d\tau = Ndt$ where $N^2=1+2A$ is lapse function.
So the four-velocity $u^{\mu}$ is $\frac{1}{N}\left(1 , v^i\right)$ where $v^i \equiv dx^i/dt$.

Now consider gauge transformations such as $\bar{t}=t+\alpha$ and $\bar{x}^i = x + \beta^{,i}$. 
Then the perturbed quantities transform as follows.
\begin{align}
    \bar{A} &= A - \dot{\alpha} \\ 
    \bar{B} &= B + \frac{\alpha}{a^2} - \dot{\beta} \\
    \bar{E} &= E - \dot{\beta} \\ 
    \bar{F} &= F - \frac{\dot{a}}{a}\alpha \\
    \bar{\delta} &= \delta + 3\frac{\dot{a}}{a}\alpha \\
    \bar{v}^i  &= v^i + \dot{\beta}^{,i}
\end{align}
Using the above transformations, the following gauge invariant quantities are defined as usual.
\begin{align}
    \Phi &\equiv A + a^2\dot{B} + 2a\dot{a}B - a^2\ddot{E} - 2a\dot{a}\dot{E} \\
    \Psi &\equiv F + a\dot{a}B - a\dot{a}\dot{E} \\
    \varDelta &\equiv \delta - 3a\dot{a}B + 3a\dot{a}\dot{E} \\
    V^i &\equiv v^i + \dot{E}^{,i}
\end{align}
These represent the gauge-invariant potential, density fluctuations, and velocity, respectively.
Using these quantities, we obtain gauge-invariant linearized Einstein equations as follows:
\begin{gather}
    \left(\frac{\dot{a}}{a}\right)^2 - 2\left(\frac{\dot{a}}{a}\right)^2\Phi + 2\frac{\dot{a}}{a}\dot{\Psi} - \frac{2\Delta\Psi}{3a^2} = \frac{8\pi G}{3}\rho_b\left(1 + \varDelta\right) \label{00inv}\\
    \frac{\dot{a}}{a}\Delta\Phi - \Delta\dot{\Psi} = -4\pi G\rho_ba^2V^i_{,i} \label{0iinv}\\
    \left[-3\frac{\ddot{a}}{a}\ + \ddot{\Psi} + 6\frac{\dot{a}}{a}\dot{\Psi} - \frac{\dot{a}}{a}\dot{\Phi} - 3\left(\frac{\dot{a}}{a}\right)^2\Phi - \frac{\Delta\Psi}{a^2}\right]\delta^i_j + \frac{1}{a^2}\left(\Phi + \Psi\right)^{,i}_{,j} \notag \\ = 4\pi G\rho_b\left(1 + \varDelta\right)\delta^i_j \label{ijinv}
\end{gather}
From $i\neq j$ of the Eq.(\ref{ijinv}), we have $\Phi + \Psi = 0$.
Furthermore, from the first order of perturbation in equations $(\ref{00inv})$ and $(\ref{ijinv})$, we can lead to the equation for $\Psi$ as
\begin{equation}
    \ddot{\Psi} + 4\frac{\dot{a}}{a}\dot{\Psi} = 0,
\end{equation}
the growing solution corresponds to $\dot{\Psi} = 0$.
We only consider the growing solution henceforth.

The conservation of the energy-momentum tensor $(\nabla_{\mu}T^{\mu}_{~\nu}=0)$ is given in the linearized order as
\begin{gather}
    \dot{\varDelta} + V^i_{,i} = 0 \label{0ec}\\
    \dot{V}_i + 2\frac{\dot{a}}{a}V_i + \frac{\Phi_{,i}}{a^2} = 0 \label{iec}
\end{gather}
From Eq.(\ref{0iinv}),(\ref{0ec}) and $a=\left(\frac{9}{4}H_0^2\right)^{1/3}t^{2/3}$, we obtain
\begin{equation}
    \Delta\Phi = 4\pi G\rho_ba^2\varDelta \label{pe}
\end{equation}
Using eq.(\ref{0ec})$\sim$(\ref{pe}), we obtain the second-order differentiation equation for the gauge-invariant density contrast $\varDelta$.
\begin{equation}
    \ddot{\varDelta} + 2\frac{\dot{a}}{a}\dot{\varDelta} - 4\pi G\rho_b\varDelta = 0
\end{equation}

\section{Spatial averaging over local domain}
We now consider the local domain large enough for the averaging meaningful but small enough compared with the horizon scale.
For example the domain with several hundred Mega parsec scale applies.
Since we approximate the domain as a homogeneous and isotropic local universe, namely we describe our local domain by Friedman equations with appropriately defined scale factor $a_D$, we expect that the (non-relativistic) density $\langle\rho\rangle_D$ defined in region $D$ obeys the following equation.
\begin{equation}
    \frac{d}{dt}\langle\rho\rangle_D + 3\frac{\dot{a}_D}{a_D}\langle\rho\rangle_D = 0
\label{rhoD}
\end{equation}
We regard that this equation defines the local Hubble parameter $H_D = \frac{\dot{a}_D}{a_D}$.

The locally averaged Friedman universe is obtained as the following step.
First we consider the spatial average of equations (\ref{00inv}) and (\ref{ijinv}).
\begin{gather}
    \left(\frac{\dot{a}}{a}\right)^2 + \frac{2\langle\Delta\Phi\rangle_D}{3a^2} = \frac{8\pi G}{3}\rho_b\left(1 + \langle\varDelta\rangle_D + 2\langle\Phi\rangle_D\right) \label{00ave} \\
    \frac{\ddot{a}}{a} - \frac{\langle\Delta\Phi\rangle_D}{3a^2} = -\frac{4\pi G}{3}\rho_b\left(1 + \langle\varDelta\rangle_D + 2\langle\Phi\rangle_D\right) \label{ijave}
\end{gather}
From the above equation, we expect the right-hand side would be the density of the local region $D$.
That is, the density $\langle\rho\rangle_D$, defined below, is considered satisfy the equation \eqref{rhoD}.
\begin{equation}
\langle\rho\rangle_D\equiv\rho_b\left(1+\langle\varDelta\rangle_D + 2\langle\Phi\rangle_D\right)
\end{equation}
Then by taking the time derivative of the averaged $00$ component Eq.(\ref{00ave}) combined with the averaged equation of the spatial component Eq.(\ref{ijave}), we obtain
\begin{equation}
    \frac{\dot{a}_D}{a_D} = \frac{\dot{a}}{a}\left(1 - \frac{2a\langle\Delta\Phi\rangle_D}{9H_0^2}\right) = \frac{\dot{a}}{a}\left(1 - \frac{1}{3}\langle\varDelta\rangle_D\right)
\end{equation}
where we have used
\begin{equation}
    \langle\Delta\Phi\rangle_D = 4\pi G \rho_b a^2\langle\varDelta\rangle_D, \quad \frac{2a}{3H_0^2} = \frac{1}{4\pi G\rho_ba^2}
\end{equation}
This equation also shows that the potential $\Phi$ is much smaller than the density contrast $\varDelta$, the result is essentially same with the previous result.
The Friedman equation may be expressed as
\begin{equation}
    \left(\frac{\dot{a}_D}{a_D}\right)^2 + \frac{10\langle\Delta\Phi\rangle_D}{9a_D^2} = \frac{8\pi G}{3}\langle\rho\rangle_D
\end{equation}

\section{Conclusion and Discussion}
Motivated by the Hubble tension, there have been many studies on the possible  resolutions.
One of them is the local variation of the cosmological parameters due to inhomogeneous matter distribution.
We have also studied the inhomogeneous universe by spatial averaging and obtained an interesting results on the relation between the local and global Hubble parameters which might explain the Hubble tension.

However the question of the gauge invariance of the result is not fully understood.
In this paper we address this question.
We employ the gauge-invariant linear cosmological perturbation theory to show that the spatial averaging is gauge invariant, namely the averaged Friedmann equation and the relation between the local and global cosmological parameters take the same form.

It is some interest to develop this treatment to the second order since the density contrast report by the observation of the K-band luminosity density is of the order $-0.5$.
Although we gave an argument based on the order of magnitude discussion of the cosmological Poisson equation, it is clearly not sufficient.
Another direction of this study is to consider  the possible interpretation by the inhomogeneity of the observation of m-z relation of Type Ia supernovae and CMB Power spectrum.
We hope to study this possibility in future.


\end{document}